\documentclass[aps,prb,a4paper,10pt,twocolumn,amsmath,showpacs,longbibliography,superscriptaddress,bibnotes]{revtex4-2} 
\usepackage{bm}
\usepackage{float}
\usepackage{dcolumn,graphicx,color,booktabs,microtype,afterpage}
\usepackage{amssymb}
\usepackage{amsmath}
\usepackage[charter,greekuppercase=italicized]{mathdesign}
\usepackage{sidecap}
\usepackage[mathlines]{lineno}
\graphicspath{{./}{figure/}}
\DeclareGraphicsExtensions{.eps,.png,.pdf,.jpg}

\renewcommand{\tablename}{Table}
\makeatletter\renewcommand{\fnum@figure}[1]{\figurename~\thefigure.~}\makeatother
\makeatletter\renewcommand{\fnum@table}[1]{\tablename~\thetable.}\makeatother

\newcount\hh \newcount\mm
\hh=\time \divide\hh by 60
\mm=\hh \multiply\mm by 60 \mm=-\mm
\advance\mm by \time
\def\now{\number\hh:\ifnum\mm<10{}0\fi\number\mm}

\usepackage[colorlinks,plainpages=false,linkcolor=blue,urlcolor=blue,citecolor=blue,pdfpagemode=UseNone,pdfstartview=FitBH]{hyperref}

\newcommand{\tcr}{\textcolor{black}}

\begin{document}
\title{Complex magnetic phase diagram in noncentrosymmetric EuPtAs}
\author{W. Xie}
\email{Present address: Deutsches Elektronen-Synchrotron (DESY),  Notkestrasse 85, 22607 Hamburg, Germany}
\affiliation{Center for Correlated Matter and Department of Physics, Zhejiang University, Hangzhou, 310058, China}
\author{S. S. Luo}
\affiliation{Center for Correlated Matter and Department of Physics, Zhejiang University, Hangzhou, 310058, China}
\author{H. Su}
\affiliation{Center for Correlated Matter and Department of Physics, Zhejiang University, Hangzhou, 310058, China}
\author{X. Y. Zheng}
\affiliation{Center for Correlated Matter and Department of Physics, Zhejiang University, Hangzhou, 310058, China}
\author{Z. Y. Nie}
\affiliation{Center for Correlated Matter and Department of Physics, Zhejiang University, Hangzhou, 310058, China}
\author{M. Smidman}
\email{msmidman@zju.edu.cn}
\affiliation{Center for Correlated Matter and Department of Physics, Zhejiang University, Hangzhou, 310058, China}
\author{T. Takabatake}
\affiliation{Center for Correlated Matter and Department of Physics, Zhejiang University, Hangzhou, 310058, China}
\affiliation{Department of Quantum Matter, Graduate School of Advanced Science and Engineering, Hiroshima University, Higashi-Hiroshima 739-8530, Japan}
\author{H. Q. Yuan}
\email{hqyuan@zju.edu.cn}
\affiliation{Center for Correlated Matter and Department of Physics, Zhejiang University, Hangzhou, 310058, China}
\affiliation{Collaborative Innovation Center of Advanced Microstructures, Nanjing 210093, China}
\affiliation{State Key Laboratory of Silicon Materials, Zhejiang University, Hangzhou 310058, China}
\date{\today}

\begin{abstract}
We report the observation of multiple magnetic transitions in single crystals of EuPtAs using magnetization, transport, and thermodynamic measurements. EuPtAs crystallizes in the noncentrosymmetric tetragonal LaPtSi-type structure (space group \emph{I}4$_1$\emph{md}) and undergoes a second-order antiferromagnetic transition at 14.8~K, which is followed by a first-order transition at about 7.5~K. Both transitions are suppressed by magnetic fields applied parallel and perpendicular to the $\emph{c}$~axis, and multiple field-induced transitions are observed for both field directions, leading to complex temperature-field diagrams with a dome-like magnetic phase for fields parallel to the $\emph{c}$~axis. Moreover, the dependence of the low temperature resistivity on the application of a training field, but the lack of remanent magnetization, suggests the presence of multiple antiferromagnetic domains in zero-field.

\begin{description}
\item[PACS number(s)]

\end{description}
\end{abstract}

\maketitle

\section{Introduction}  
Magnets with a noncentrosymmetric structure often exhibit complex magnetic structures due to the presence of Dzyaloshinskii-Moriya (DM) interaction \cite{I.Dzaloshinsky1958, Moriya1960}.  Unlike other exchange interactions which typically favor collinear spin structures, the DM interaction tends to orientate the spins perpendicular to each other. 
The competition between these competing tendencies can lead to noncollinear magnetic structures which could be topologically nontrivial in real space \cite{2013Nagaosa}.   
For example, a magnetic skyrmion phase with a topologically nontrivial spin texture has been experimentally discovered in MnSi, which has the chiral B20-type structure (space group \emph{P}2$_1$3) \cite{MnSi1976,MnSi2009,MnSi2011}. Subsequently, a series of additional compounds with the B20 structure have also been identified to host magnetic skyrmions, such as Fe$_{1-x}$Co$_x$Si \cite{1FeCoSi2010,2FeCoSi2010}, FeGe \cite{FeGe2008,FeGe2011}, and Cu$_2$OSeO$_3$ \cite{1CuO2SeO3,2CuO2SeO3}. 
Recently EuPtSi, which also crystallizes in a chiral cubic structure (LaIrSi-type, space group \emph{P}2$_1$3) \cite{Si1986structure},  was shown to be an unusual example of an \emph{f}-electron system hosting magnetic skyrmions \cite{2017hall,2018neutron, EPS2021}, where the size of single skyrmion for $H \parallel$ [111] is only 18 $\AA$, ten times smaller than those in MnSi \cite{EPS_size}.  
In addition to these compounds with chiral structures, it was found that compounds with a noncentrosymmetric polar structure are also promising candidates for realizing skyrmions, such as GaV$_4$S$_8$ with a lacunar spinel structure \cite{2015GaV4S8}, which hosts a N\'{e}el-type skyrmion phase. For CeAlGe with a polar tetragonal structure, evidence for a meron/antimeron phase has been reported, which has a double-\emph{k} spin structure with a topological charge \emph{Q} = 1/2 \cite{2020CeAlGe}, in contrast to a skyrmion lattice which has a triple-\emph{k} structure with \emph{Q} = 1.

Although quite a few types of materials have been found to host skyrmion-like topological spin textures, such explorations in rare-earth compounds are still relatively limited after the report of EuPtSi, with the focus on some Gd-based centrosymmetric compounds where magnetic skyrmions with reduced size were commonly revealed in the absence of the DM interaction \cite{Gd2PdSi3, GdRu2Si2, Gd3Ru4Al12, Gd_PRL}. Here we focus on the Eu-based compound EuPtAs. Unlike other reported members in the Eu\textit{TX} series \cite{2015review-RTX}, EuPtAs crystallizes in the polar tetragonal LaPtSi-type structure (space group \emph{I}4$\rm_1$\emph{md}) \cite{As1986structure} that is isostructural to CeAlGe. The crystal structure is displayed in the inset of Fig.~\ref{fig1}. For the Eu atoms, the nearest neighbor (NN) distance of 4.23 ${\AA}$ (in plane) is very close to the next nearest neighbor (NNN) distance 4.28 ${\AA}$ (out of plane), which are also comparable to the Ce-Ce distances in CeAlGe. However, in comparison to Ce$^{3+}$ in CeAlGe, the Hund's rule ground state of Eu$^{2+}$ in EuPtAs has a total orbital angular momentum \emph{L} = 0, which offers the opportunity to examine the magnetic properties in the same crystal structure but with greatly reduced magnetocrystalline anisotropy. On the other hand, the magnetic atoms in both compounds form nearly equilateral triangular sublattices where magnetic frustration is possible, which has been proposed to be another important mechanism for the realization of magnetic skyrmions, even in centrosymmetric compounds \cite{Gd2PdSi3}.

In this paper, we report magnetization, transport, thermodynamic measurements of EuPtAs single crystals grown by a Pb-flux method, revealing that EuPtAs undergoes two magnetic transitions in zero-field at 14.8 K and 7.5 K, which were both continuously suppressed in applied magnetic fields up to 9 T.
We construct the magnetic field-temperature (\emph{H}--\emph{T}) phase diagrams for magnetic fields applied parallel and perpendicular to the easy $\emph{c}$~axis, featuring a number of field-induced magnetic phases. 

\section{Experimental methods}
\begin{figure}[!htb]
     \begin{center}
     \includegraphics[width=0.95\columnwidth]{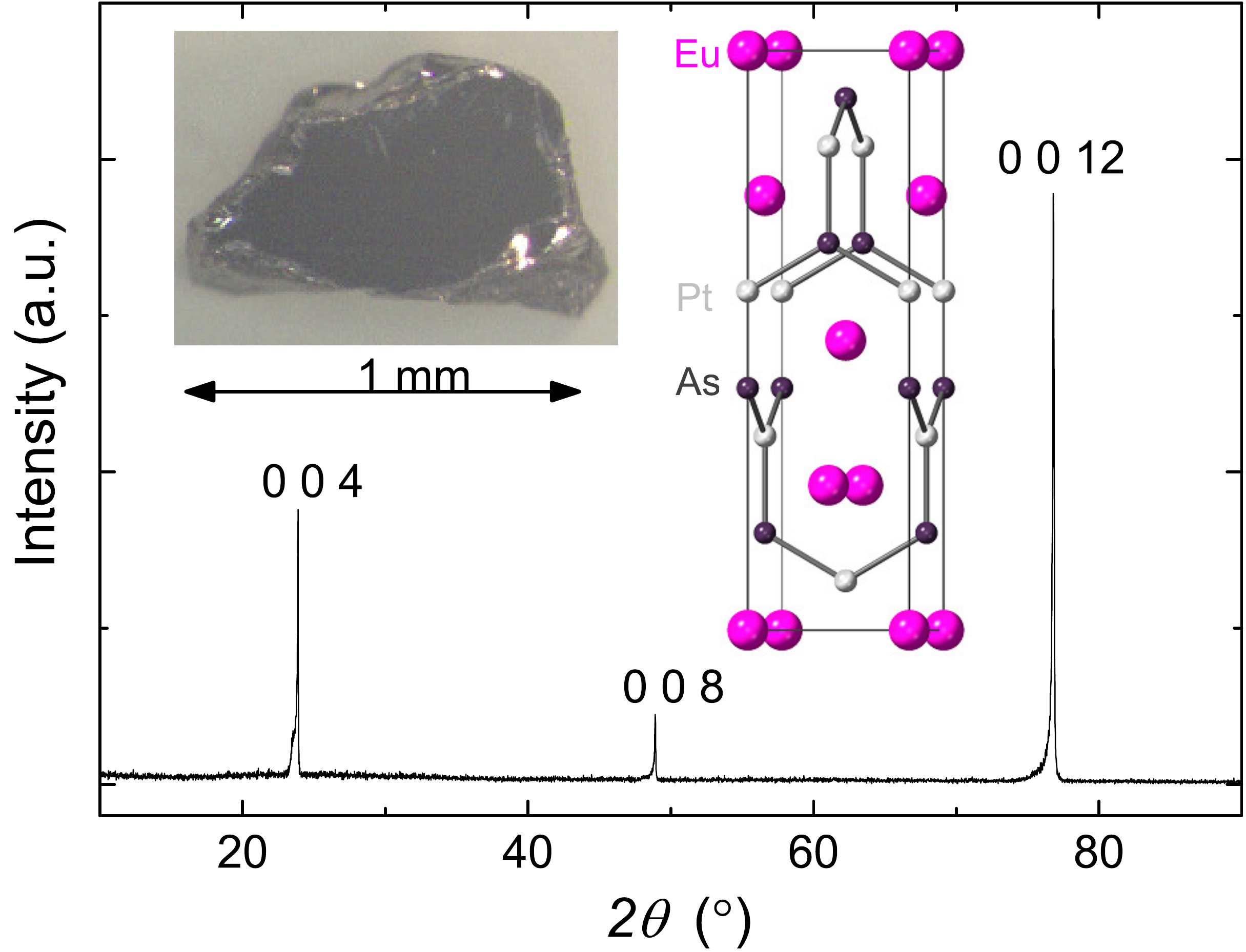}
     \end{center}
     \caption{(Color online) XRD pattern of a EuPtAs single crystal with a surface plane perpendicular to the tetragonal \emph{c} axis. The left inset shows a photograph of a piece of single crystal with a typical length of 1.0 mm. The unit cell of EuPtAs is shown in the right inset. }
     \label{fig1}
\end{figure}

Single crystals of EuPtAs were grown using a Pb-flux method.
 Eu pieces (99.9\%), Pt powder ($99.9+\%$), and As grains ($99.999\%$) from Alfa Aesar were mixed in an off-stoichiometric ratio,
and put into an alumina crucible together with excessive Pb ingots ($99.99+\%$). The crucible was sealed in an evacuated quartz ampule,
which was slowly heated  to 1100$^\circ$C and kept there for 2 days before being cooled at a rate of 3$^\circ$C/h to 750$^\circ$C, at which temperature the ampule was removed from the furnace and centrifuged.
Shiny plate-like crystals with typical dimensions of
$1\times 0.5\times0.2 \textrm{mm$^3$}$ were obtained.
In order to remove the residual Pb flux, the crystals were etched in dilute hydrochloric acid and polished.
A photograph of a piece of single crystal is shown in the inset of Fig.~\ref{fig1}.
The crystal structure and orientation were characterized by x-ray diffraction (XRD) using a PANalytical
X'Pert MRD diffractometer with Cu K$_{\alpha1}$ radiation monochromated by
graphite. The chemical composition was
determined by energy-dispersive x-ray spectroscopy (EDS) using a Hitachi SU-8010 field emission scanning electron microscope.
The electrical resistivity and heat capacity were
measured using a Quantum Design Physical Property Measurement System (QD PPMS-9T) with a $^3$He insert.
The magnetic susceptibility and magnetization were measured down to 2 K using a Quantum Design Magnetic Property Measurement System (QD MPMS-5T)
and a vibrating sample magnetometer (VSM) on a Physical Property Measurement System (PPMS-14T).

\begin{figure}[!htb]
\begin{center}
     \includegraphics[width=0.95\columnwidth]{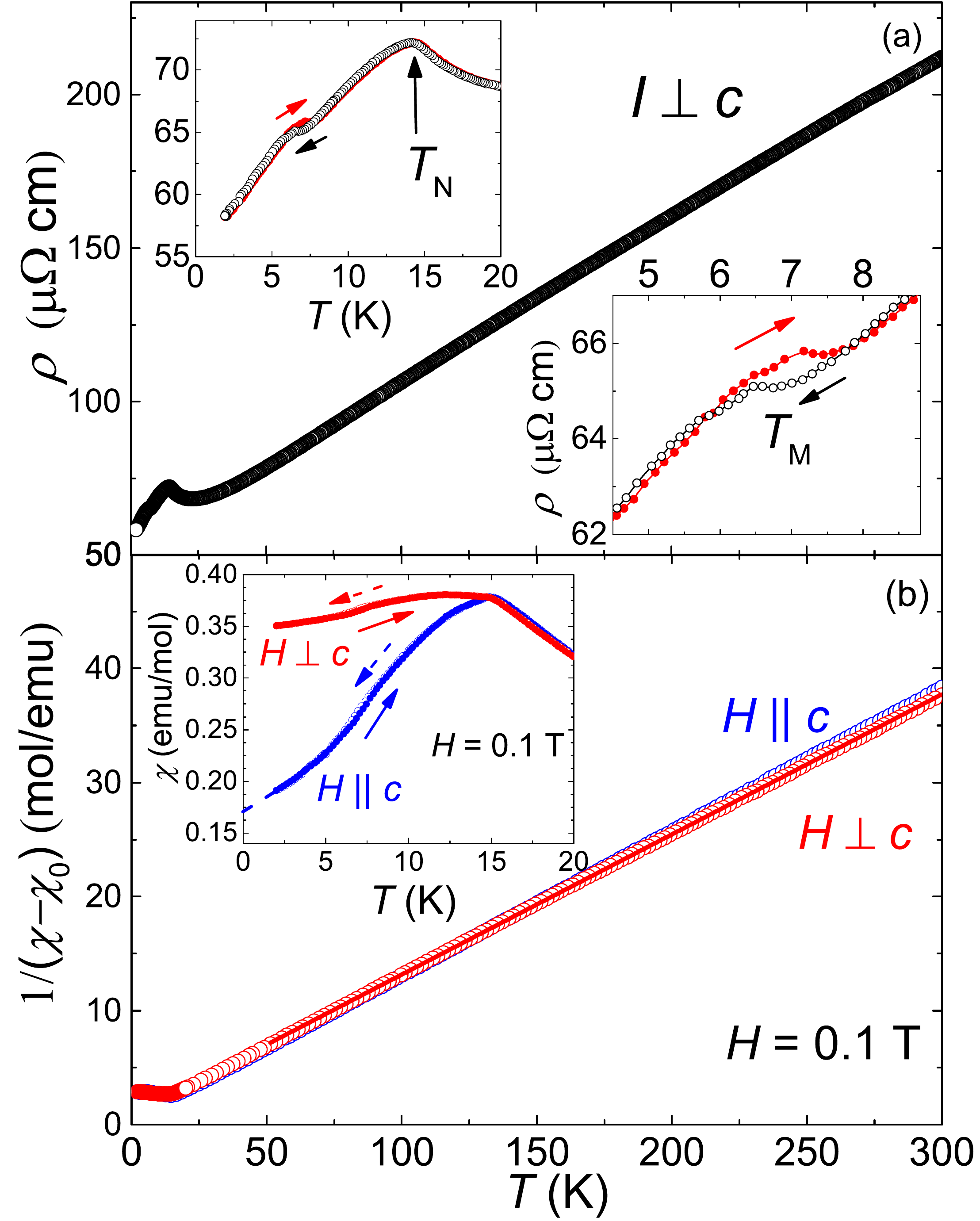}
     \end{center}
     \caption{(Color online) (a) Temperature dependence of the resistivity $\rho$(\emph{T}) in the temperature range 300--1.9 K. The insets display low-temperature enlargements in order to show the transitions at \emph{T}$\rm_N$ and \emph{T}$\rm_M$. (b) Inverse magnetic susceptibility from 300 K to 2K for both \emph{H} $\parallel$ \emph{c} and \emph{H} $\perp$ \emph{c} measured in an applied field of 0.3 T. The solid lines are fits of the high-temperature data with the modified Curie-Weiss law. The inset shows the low temperature part of $\chi(T)$ for both directions.}
\label{fig2}
\end{figure}

\section{Results and discussions}

\subsection{Zero-field characterization}

Figure~\ref{fig1}(a) shows the XRD pattern measured on a piece of plate-shaped single crystal, in which the diffraction peak positions match well with the expected (00$l$)($l$=integer) peak positions for EuPtAs \cite{As1986structure}, indicating that the plate-like surface is oriented perpendicular to the \emph{c} axis. The EDS measurements performed at multiple positions on the crystal show an average atomic ratio of 1 : 1.02 : 1.01 for Eu:Pt:As, which is very close to the stoichiometric EuPtAs composition.

The temperature dependence of the electric resistivity $\rho(T)$, measured with a current $I \perp c$, is shown in  the main panel of Fig.~\ref{fig2}(a). The nearly linear behavior in $\rho(T)$ is commonly observed for Eu$^{2+}$-based intermetallics \cite{2001EuPtSn,EuTGe3}, which is generally ascribed to arising from the electron-phonon scattering \cite{EuTGe3, Bloch-Gruneisen}. At low temperature, $\rho(T)$ shows a peak at the magnetic ordering temperature \emph{T}$\rm_N$ $\approx$ 14.8 K and a cusp at \emph{T}$\rm_M$ = 6.5 K upon cooling (7 K upon warming). While negligible differences are found at \emph{T}$\rm_N$ between the cooling and warming processes, hysteresis is observed at \emph{T}$\rm_M$, as shown in the insets of Fig.~\ref{fig1}(a). 

The temperature dependence of the magnetic susceptibility $\chi(T)$ is displayed in Fig.~\ref{fig2}(b), \tcr{where the high temperature behavior can be well described by the modified Curie-Weiss form $\chi = \chi{_0} + C/(T-\theta_P)$ with an effective magnetic moment of $\mu$$_{eff}$ = 7.95$\mu$$_B$/Eu and a Weiss temperature $\theta_P$ = --4.7 K for $H\parallel c$, while for \emph{H} $\perp c$, $\mu$$_{eff}$ = 8.03$\mu_B$/Eu and $\theta_P$ = --7.3 K. For both directions there is a temperature-independent susceptibility $\chi{_0} =$ 0.004 emu/mol.} The magnitude of $\mu$$_{eff}$ is close to the expected value of 7.94 $\mu$$_B$ for the well-localized  Eu$^{2+}$ electronic state with $J$ = 7/2. The inset of Fig.~\ref{fig2}(b) displays the low temperature $\chi(T)$ for both $H\parallel c$ and \emph{H} $\bot$ \emph{c}, where a cusp is clearly observed at $T\rm_{N}$, while a weak slope change with small thermal hysteresis occurs at $T\rm_{M}$. 
The anisotropic behavior in $\chi(T)$ at $T \leq T\rm_N$ indicates an antiferromagnetic (AFM) transition with the \emph{c}-axis as the easy axis. In addition, extrapolation of the \emph{c}~axis susceptibility  $\chi_c(T)$ down to zero temperature gives a finite $\chi_c$(0) of about 0.168 emu/mol, yielding the ratio $\chi_c$(0)/$\chi$(\emph{T}$\rm_N$) = 0.44. This feature is inconsistent with a collinear AFM structure of the Eu spins along the \emph{c} axis, for which $\chi_c$(0) should be zero \cite{2016EuMn2As2}. A possible explanation is that the Eu spins tilt away from the \emph{c}~axis, forming a noncollinear magnetic structure. 
Two consecutive magnetic transitions at low temperature are also observed in some other Eu-based intermetallics, for instance, EuPt$_2$As$_2$ \cite{EuPt2As2} and EuPd$_2$As$_2$ \cite{EuPd2As2}. While the spin structure change in EuPt$_2$As$_2$ is still not well-understood, the spins in EuPd$_2$As$_2$ are reported to be mainly orientated in the \emph{ab}~plane below $T\rm_{N}$, which cant towards the \emph{c}~axis upon crossing $T\rm_{M}$. Consequently, $\chi_c$ is flat between $T\rm_N$ and $T\rm_M$ and then decreases significantly below $T\rm_M$, while $\chi_{ab}$ decreases continuously below $T\rm_N$ with a smaller decreasing rate upon crossing $T\rm_M$ \cite{EuPd2As2}.  However,  both $\chi_c(T)$ and $\chi_{ab}(T)$ for EuPtAs decrease monotonically below $T\rm_N$ where $\chi_c(T)$ displays a steeper drop. In addition, the rate at which both $\chi_c(T)$ and $\chi_{ab}(T)$ decrease varies continuously with temperature, where there is an inflection point near $T\rm_{M}$. This is in contrast to that observed in EuPt$_2$As$_2$ \cite{EuPt2As2} and EuPd$_2$As$_2$ \cite{EuPd2As2}, which might be due to a \emph{gradual} spin canting with decreasing temperature.

 \begin{figure}[!htb]
\centering\includegraphics[width=\columnwidth]{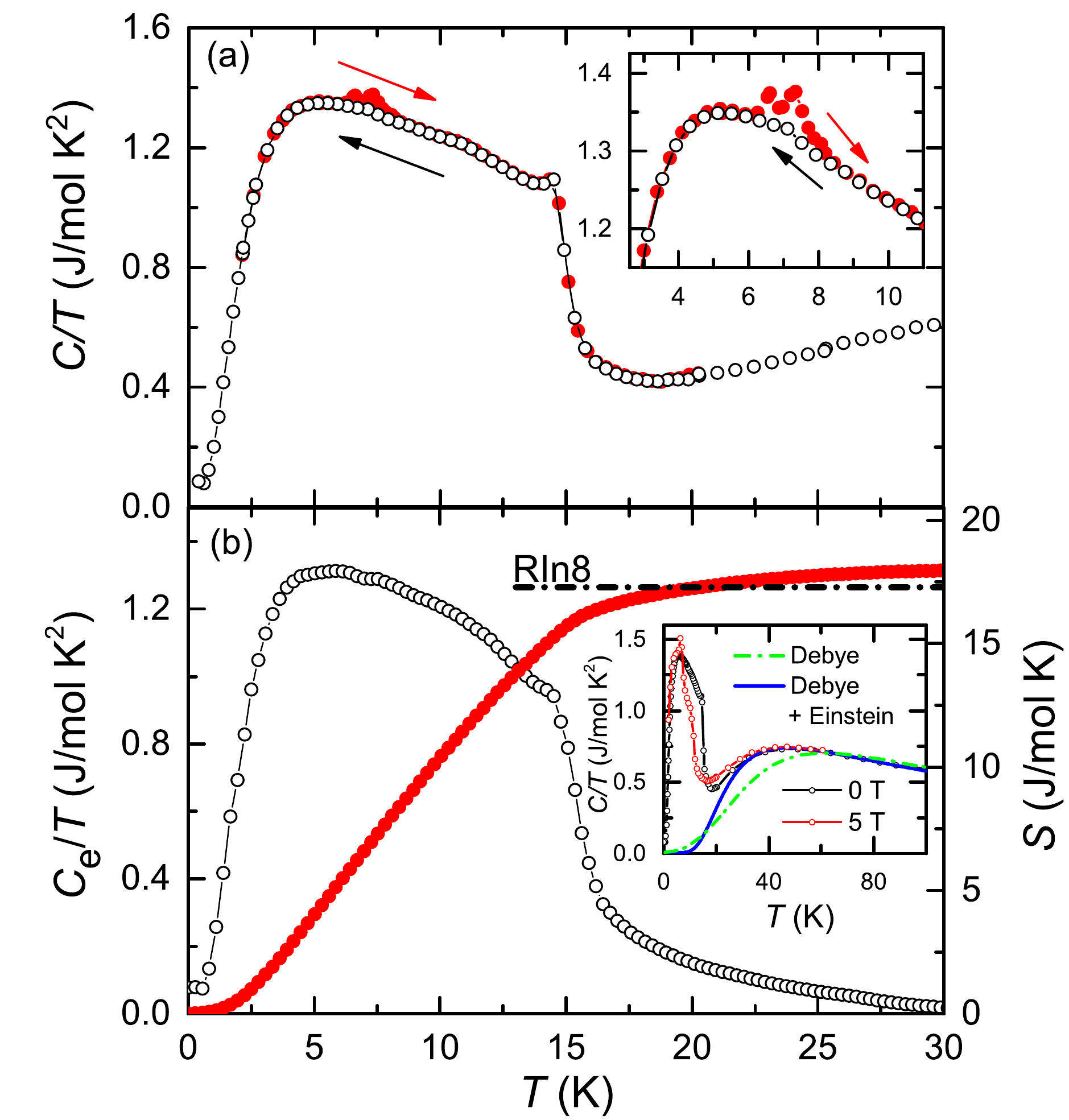}
\caption{(Color online)(a) Low-temperature specific heat as \emph{C}(\emph{T})/\emph{T} measured upon warming and cooling, as indicated by the arrows. The inset enlarges the data to display the first order transition at \emph{T}$\rm_M$. (b) The magnetic entropy calculated (right-hand axis) by integrating the electronic contribution \emph{C}$_\mathrm{e}$(\emph{T})/\emph{T} (left-hand axis). The inset shows the fits to the high-temperature data of \emph{C}(\emph{T})/\emph{T} for two adopted models.  }
\label{fig3}
\end{figure}

\begin{figure}[!htb]
\centering\includegraphics[width=0.95\columnwidth]{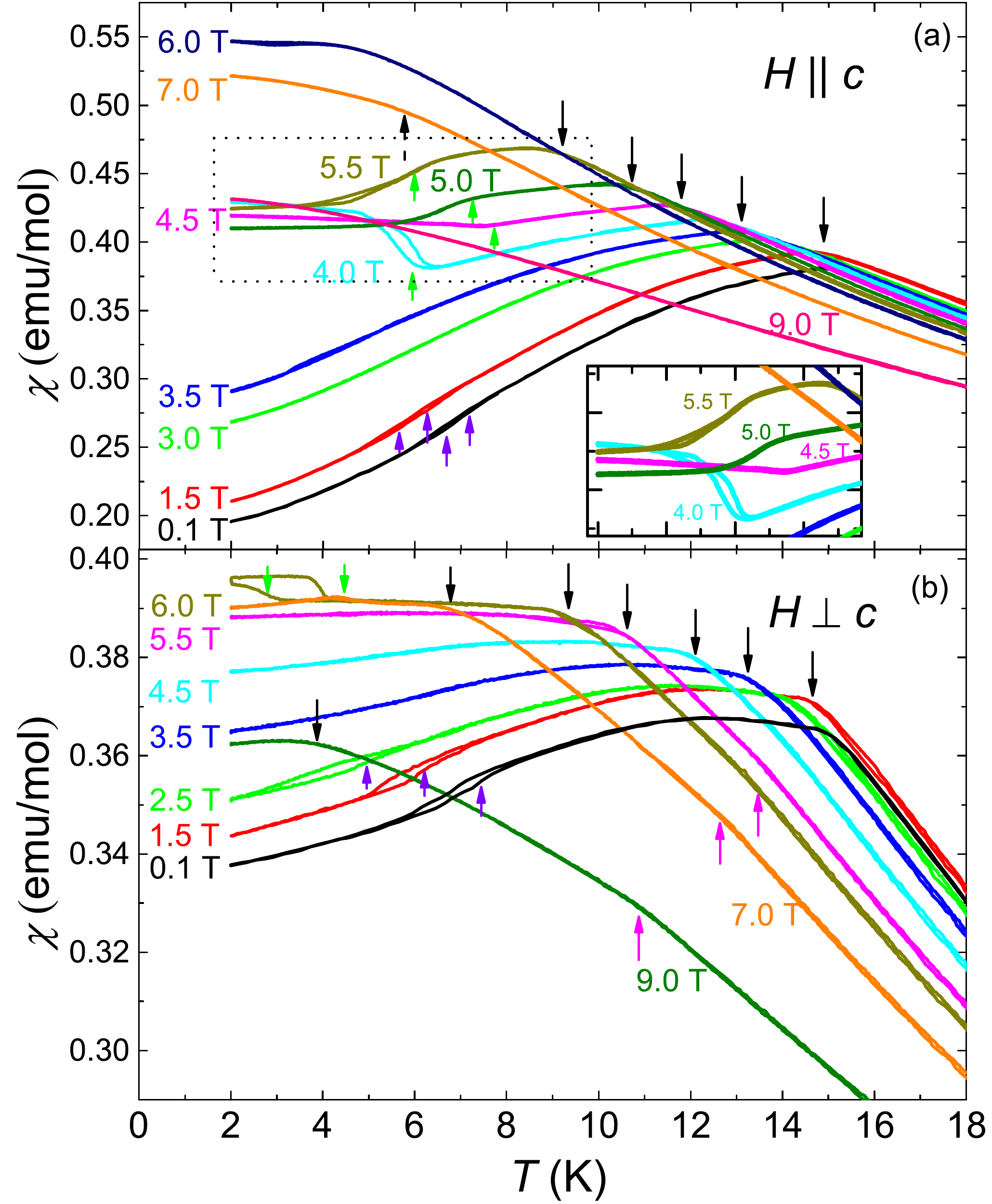}
\caption{(Color online)  Low temperature magnetic susceptibility $\chi$(\emph{T}) in magnetic fields up to 9 T applied (a) parallel to, and (b) perpendicular to the \emph{c} axis. The different colored arrows show the positions of different transitions in both (a) and (b): black downward arrows denote $T\rm_N$; the violet upward denote $T\rm_M$; the green (or magenta) arrows show the field-induced transitions below (or above) $T\rm_N$. The inset of panel (a) enlarges the range in the dotted square. }
\label{fig4}
\end{figure}

The temperature dependence of the specific heat as $C(T)/T$ is shown in Fig.~\ref{fig3}(a). A jump at about 14.9 K and two small peaks centered at 7.3~K (upon warming) are observed. The former corresponds to the AFM transition at $T\rm_N$, while the later peaks arise due to the magnetic transition at $T\rm_M$, where the difference between warming and cooling processes is consistent with $T\rm_M$ being a first-order transition.
The magnetic entropy is shown in Fig.~\ref{fig3} (b), which was calculated by integrating the electronic contribution \emph{C}$_\mathrm{e}/T$ after subtracting the lattice contribution, where \emph{C}$_\mathrm{e}/T$ was assumed to be linear to $T$ for $T \leq $ 0.4 K. Note that due to the lack of a suitable nonmagnetic reference compound, the high temperature phonon contribution was fitted with various models. The results are shown in the inset of Fig.~\ref{fig3}(b), which indicates that a model combining both Debye and Einstein contributions [$C_L(T) = (1-p)C_D(T) + pC_E(T)$ \cite{2015EuTGe3}] can reasonably describe the experimental data at high temperatures, with fitted values of the Debye temperature $\Theta$$\rm_D$ = 406~K, Einstein temperature $\Theta\rm_E$ = 111~K, and a weighting factor $p=0.66$.  
The entropy released at \emph{T}$\rm_N$ is slightly less than R$\ln$8, the full entropy expected for Eu$^{2+}$ with an eight-fold degenerate ground state multiplet ($J$ = 7/2). The integrated magnetic entropy reaches R$\ln$8 at 20 K, and then saturates for $T \geq$ 30 K. The additional entropy above 20K likely results from an incomplete subtraction of the phonon contribution \cite{2016EuRhGe3}. A notable feature in \emph{C}$_\mathrm{mag}/T$ is a tail above $T\rm_N$. This could arise from magnetic fluctuations or short range magnetic correlations above $T\rm_N$, which seems to be further enhanced upon applying a magnetic field of 5 T in the present case, as shown in the inset of Fig.~\ref{fig3}(b).

\subsection{In-field measurements}

Figures~\ref{fig4} (a) and (b) show $\chi (T)$ for various magnetic fields applied along the $c$ axis and $ab$ plane, respectively. With increasing field, \emph{T}$\rm_N$ is suppressed to lower temperatures with the corresponding peak becoming broader for both directions. \emph{T}$\rm_M$ is also continuously suppressed with fields, and is no longer observed down to 1.9 K for \emph{H} $\geq$ 3 T. A field-induced transition [marked by the upward green arrows] is observed for \emph{H} $\geq$ 3.5 T for $H\parallel c$. $\chi(T)$ exhibits a rather linear behavior down to the lowest temperature after crossing this transition (see the inset), which displays a non-monotonic field-dependence, first shifting to higher temperatures, and then to lower temperatures with increasing field, before disappearing above 6 T. For $H \perp c$, the field-induced transition appears for \emph{H} $\geq$ 5.5 T, where there is an abrupt increase in $\chi(T)$ below the transition, together with thermal hysteresis.  In addition, anomalies are also resolved above $T\rm_N$  for $H \geq$ 3.5 T [upward magenta arrows] for $H \perp c$, which become increasingly prominent at higher fields. These behaviors in $\chi$(\emph{T}) indicate multiple field induced phases in EuPtAs.

\begin{figure}[!htb]
\centering\includegraphics[width=1.0\columnwidth]{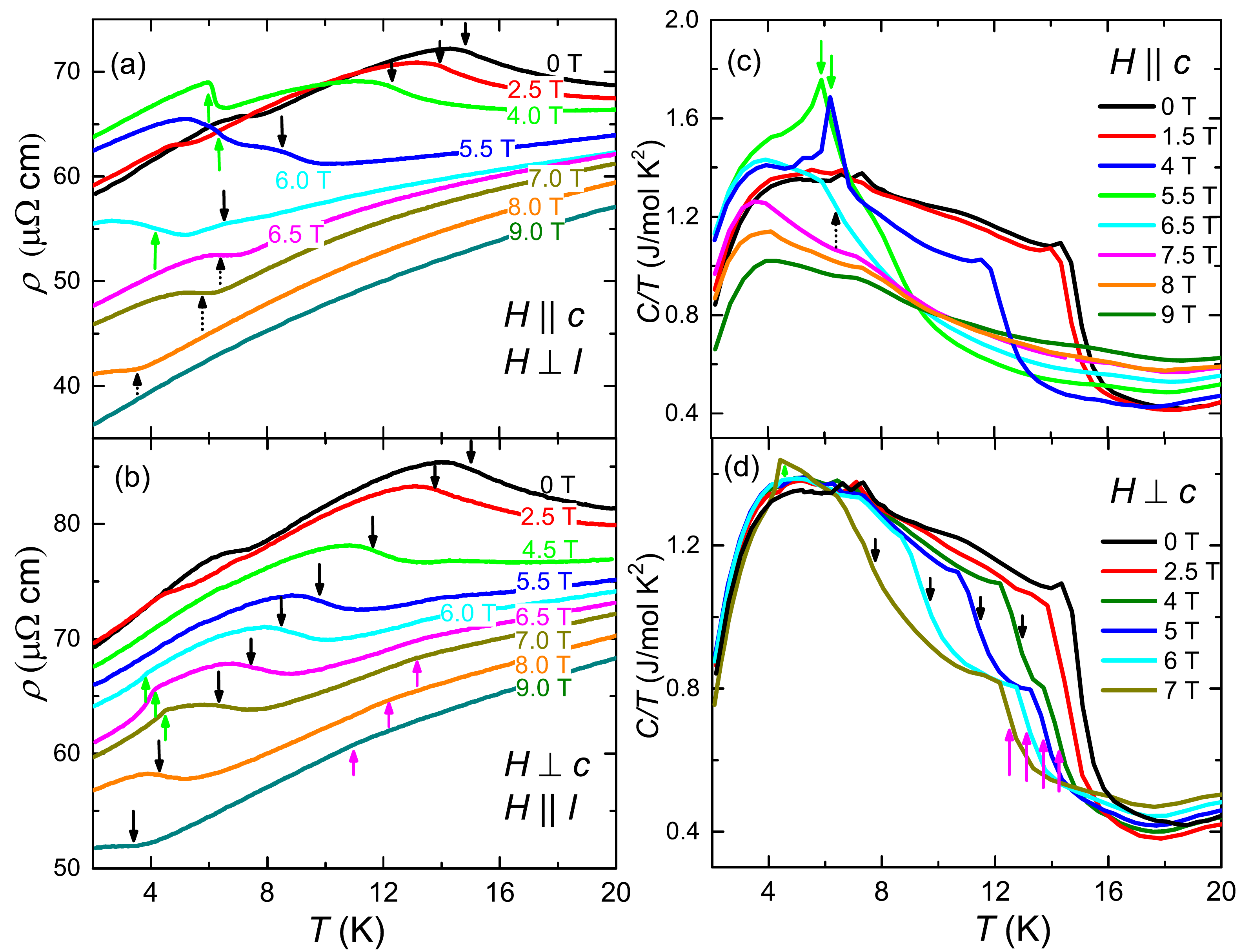}
\caption{(Color online) (a)(b) Temperature dependence of the resistivity $\rho$(\emph{T}), and (c)(d) specific heat \emph{C}(\emph{T})/\emph{T} in constant magnetic fields up to 9 T applied parallel and perpendicular to the \emph{c} axis, respectively. The arrows with different colors are used to mark different phase transitions, which are labeled in the same manner as those in Fig.~\ref{fig4}.}
\label{fig5}
\end{figure}

To further track the transitions in EuPtAs, $\rho(T)$ and \emph{C}(\emph{T}) were measured in applied magnetic fields up to 9~T. Overall, the absolute value of $\rho(T, H)$ is suppressed in applied fields, where anomalies at the various phase transitions are also detected. 
As shown in Fig.~\ref{fig5}(a), for $H \parallel c$ the broad peak at $T\rm_N$ and small cusp in $\rho(T)$ at $T\rm_M$ are both suppressed monotonically to lower temperatures, and the latter disappears at around 3 T. 
At higher fields an abrupt jump appears, and the field dependence of the anomaly is consistent with that found in $\chi (T)$ for the same field direction. For $H \geq$ 6 T, broad humps are observed in $\rho(T)$ below 8 K, as shown by the dotted upward arrow.  However, a lack of well-defined corresponding anomaly in $\chi(T)$ and $C(T)/T$ [dotted upward arrow in Fig.~\ref{fig4}(a) and Fig.~\ref{fig5}(c)] is observed, indicating that this may correspond to a crossover to the spin-polarized state. For $H \geq$ 3 T along the $c$~axis, the small peaks in $C(T)/T$ at $T\rm_M$ disappear while another sharp peak is observed (vertical green arrows), which has an asymmetric shape with a tail on the low-temperature side. This is assigned to be a field-induced transition, which disappears in fields above 6 T, being consistent with $\chi(T, H)$ and $\rho(T, H)$. As shown in Fig.~\ref{fig5}(c), all the data curves of $C(T)/T$ display a broad hump at around 3 K. This feature is frequently observed in other Eu-based compounds with divalent Eu, and is generally ascribed to the Zeeman splitting of the $J$ = 7/2 mulitplet of the Eu$^{2+}$ ions under internal and external magnetic fields \cite{EuNi5As3, EuCu2As2, EuB6}.

For \emph{H} $\perp$ \emph{c} shown in Fig.~\ref{fig5}(b), in addition to the broad peak at $T\rm_N$ and the small hump at $T\rm_M$ (suppressed for $H \geq$ 3 T), a weak but discernible slope change is observed at temperatures above $T\rm_N$ in $\rho(T)$ for $H \geq$ 4 T, the position of which slowly shifts to lower temperature with increasing field. On the other hand, a clear sharp drop is observed at lower temperatures in the field range of 6.0--7.5 T [upward green arrows], which is assigned to be another field induced transition for \emph{H} $\perp$ \emph{c}. Note that all these transitions are also manifested in $C(T)/T$, as shown in Fig.~\ref{fig5}(d). $C(T)/T$ shows an increase upon cooling prior to an abrupt jump at $T\rm_N$. This gradual increase is further enhanced under magnetic fields, which is likely due to short range correlations or fluctuations.  The overall behavior for $H \leq$ 3 T is similar to that for \emph{H} $\parallel$ \emph{c}, however for $H \geq$ 4 T,  $C(T)/T$ shows two steps upon cooling. From comparing the data of $C(T)/T$ with $\rho(T, H)$ and $\chi(T, H)$, it can be seen that the higher temperature step corresponds to a field induced transition above $T\rm_N$  [magenta arrows], while the lower temperature step is the transition at $T\rm_N$, which becomes broader in higher fields. For $H$ = 6.0--7.5 T, an asymmetric peak is observed in $C(T)/T$ [upward green arrow], which should correspond to the field induced transition below $T\rm_N$ . 

\begin{figure}[!b]
  \centering
  \includegraphics[width=1.0\columnwidth]{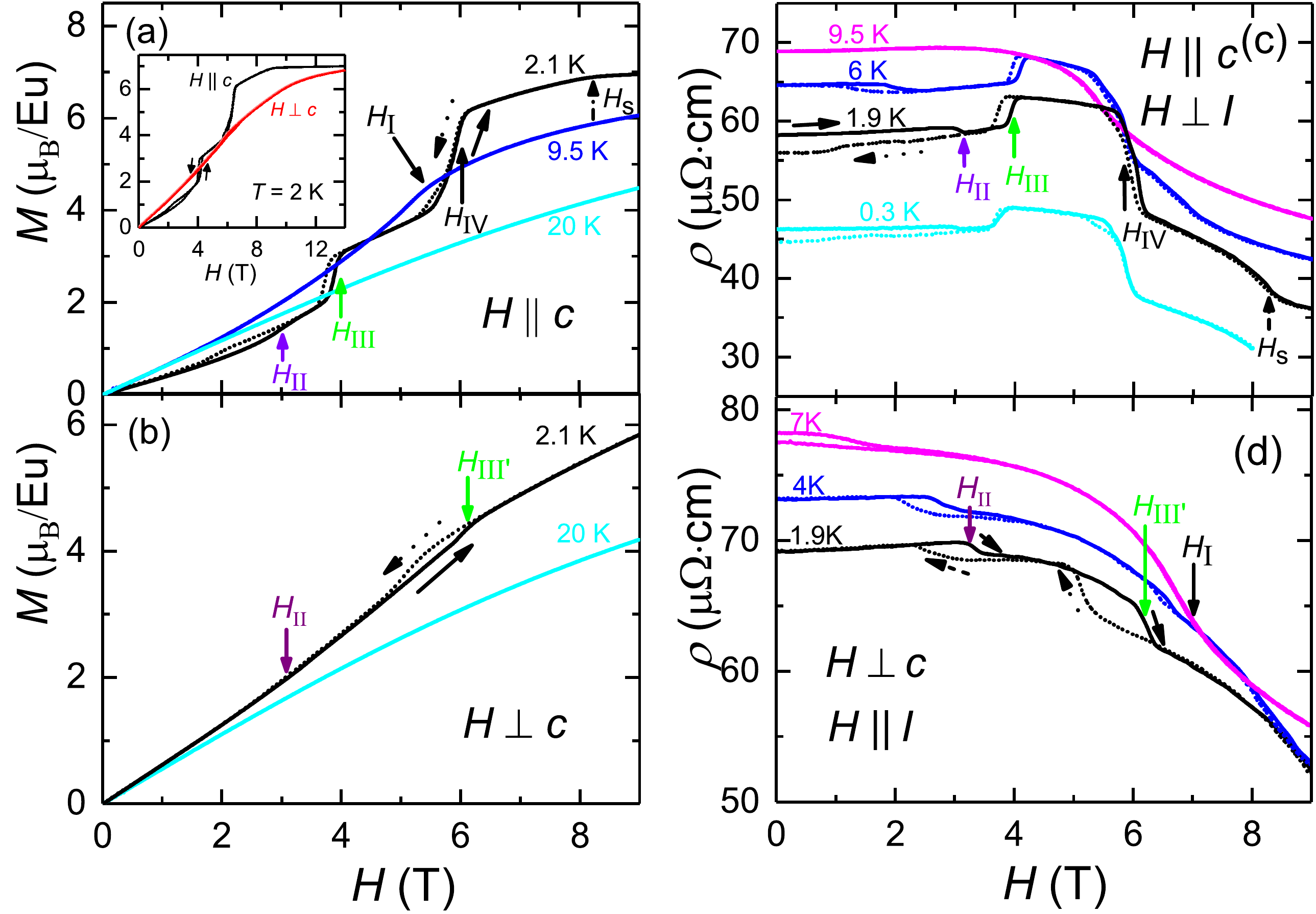}
  \caption{(Color online)  Isothermal magnetization \emph{M}(\emph{H})  measured at several temperatures for (a) \emph{H} $\parallel$ \emph{c}, and (b) \emph{H} $\perp$ \emph{c}. The field dependence of the electrical resistivity $\rho(H)$ is also displayed for (c) \emph{H} $\parallel$ \emph{c}, and (d) \emph{H} $\perp$ \emph{c}. \tcr{The inset in (a) shows the \emph{M}(\emph{H}) curves at 2 K with fields up to 14 T.} Only the $\rho(H)$ data at $T = 0.3~K$ in (c) is shifted. In all the figures, the vertical arrows show the positions of  transitions, for which the same colors are used as in Fig.~\ref{fig4} and Fig.~\ref{fig5}. The directions of the field-ramping are indicated by tilted arrows. }
\label{fig6}
\end{figure}

The isothermal magnetization \emph{M}(\emph{H}) and field dependence of the resistivity $\rho(H)$ for $H\parallel c$ and \emph{H} $\perp$ \emph{c} are displayed in Fig.~\ref{fig6}. Note that in all main panels, both the results for ramping up [0--9 T, solid lines] and ramping down processes [9--0 T, broken lines] are displayed, where the data at the lowest temperature were measured first after cooling from 300 K in zero-field, and the data at higher temperatures were subsequently measured. 
In $M(H) (H \parallel c)$, four anomalies can be resolved in the data measured upon ramping up the field at 2.1 K: a weak slope change with broad hysteresis at \emph{H}$\rm_{II}$ = 3.0~T, step-like increases at \emph{H}$\rm_{III}$ = 3.9~T and \emph{H}$\rm_{IV}$ = 6.0~T, and a kink at \emph{H}$\rm_S$ = 8.2~T, above which $M(H)$ is saturated at 7$\mu$$_B$/Eu, the saturated moment expected for fully polarized Eu$^{2+}$ spins with \emph{J} = 7/2. Near \emph{H}$\rm_{III}$, $M(H)$ jumps by approximately 1$\mu$$_B$/Eu and then linearly increases with field. The extrapolation of this linear part to \emph{H} = 0 yields a finite intercept. This is generally an indication of a spin-flip transition \cite{2001Blundell}. The same features occur at \emph{H}$\rm_{IV}$, where the magnetization jump is about 2$\mu$$_B$/Eu. At temperatures above 7.5 K but below 15 K, only one transition can be found in \emph{M}(\emph{H}) which shows a relatively broad anomaly labelled \emph{H}$\rm_{I}$ in Fig.~\ref{fig6}(a).  For $T \geq$ 15 K, $M(H)$ shows a sublinear field dependence with no detectable anomaly.
All these transitions can also be observed in the isothermal $\rho(H)(H \parallel c)$ data at the corresponding temperatures, as shown in Fig.~\ref{fig6}(c).
For  $T \leq$ 7.5 K, the $\rho(H)(H \parallel c)$ curves display a kink with a broad hysteresis loop in low fields, which then abruptly jumps (by about 4 $\mu\Omega$-cm for 1.9 K) at \emph{H}$\rm_{III}$, flattens, and subsequently drops (by about 10 $\mu\Omega$-cm for 1.9 K) at \emph{H}$\rm_{IV}$. After that, the data decreases slowly with increasing field and bends at \emph{H}$\rm_S$. At 7.5 K $\leq T \leq$ 15 K, $\rho(H)(H \parallel c)$ exhibits a broad drop at fields corresponding to \emph{H}$\rm_{I}$. 

For \emph{H} $\perp$ \emph{c}, two transitions are observed in $M(H)$ at 2 K: one at 3 T [labelled as $H\rm_{II}$] showing a weak slope change, and another at 6.0 T [labelled as $H\rm_{III'}$] showing a sharp change of slope together with a hysteresis extending from 2.5 T to 6.1 T. Consistently, $\rho(H)(H \perp c)$ [Fig.~\ref{fig6}(d)] displays two step-like decreases at $H\rm_{II}$ and $H\rm_{III'}$  with sizable hysteresis loops. 
With increasing temperature, $H\rm_{II}$ ($H\rm_{III'}$) shifts to lower (higher) fields. Above 5 K, $\rho(H)(H \perp c)$ displays a single broad drop (labelled as $H\rm_{I}$), at around 7 T for $T = 7$ K. It can be seen below that the temperature dependence of $H\rm_{I}$ and $H\rm_{II}$ correspond to the field dependence of $T\rm_N$ and $T\rm_M$, respectively.  \tcr{Magnetization measurements have been further performed up to 14 T at 2 K. 
As shown in the inset of Fig.~\ref{fig6}(a), the $M(H)$ curves show no additional anomalies at higher fields for either field direction. Note that for fields applied parallel to the $c$-axis, the magnetization saturates at \emph{H}$\rm_S$ with a moment of 7$\mu$$_B$/Eu, while perpendicular to the $c$-axis there is no full saturation even up to 14 T.}

\begin{figure}[!htb]
\centering\includegraphics[width=0.9\columnwidth]{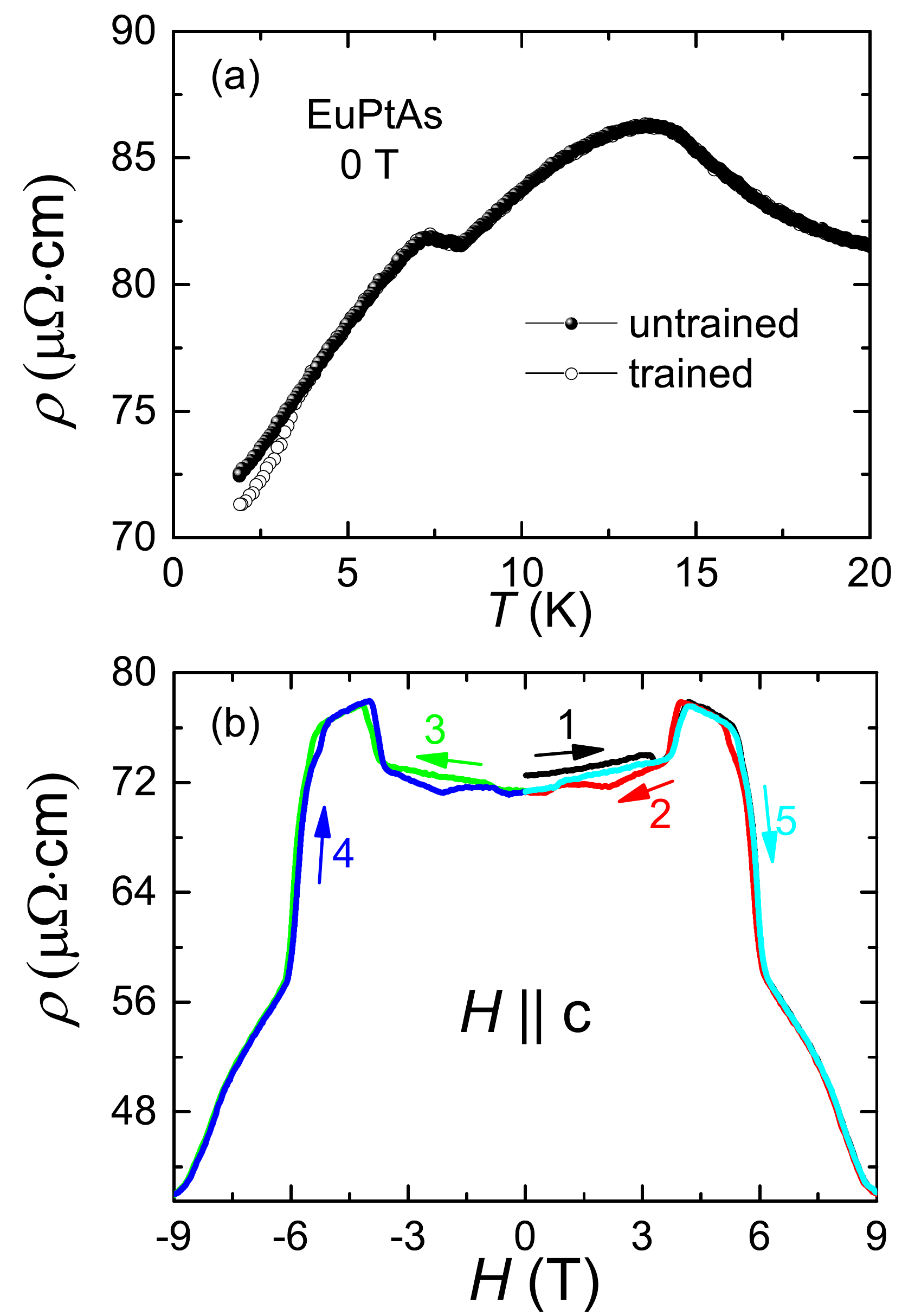}
\caption{(Color online) The effects of cooling history on the resistivity: (a) a comparison of $\rho(T)$ at zero-field for \emph{trained} and \emph{untrained} cases, as described in the text. (b) $\rho(H)$ measured after cooling from 300 K in zero field. The numbers and arrows denote the order in which the measurements were performed, and the directions of the field scans, respectively.}
\label{fig7}
\end{figure}

Note that at 0.3 K and 1.9 K, the broad hysteresis loop in $\rho(H)(H \parallel c)$ does not close when the field is reduced down to 0 T, while no zero-field hysteresis or remnant magnetization is observed in $M(H)(H \parallel c)$ in Fig.~\ref{fig6}(a).  
We further found that the hysteresis near zero field is only observed in $\rho(H)(H \parallel c)$ between the initial field scan after zero-field cooling, and the subsequent field-sweeps, as shown in Fig.~\ref{fig7}(b). On the other hand, the hysteresis at zero field also disappears if the sample is cooled in a field of 9 T and then measured upon ramping down to zero, and back up to 9 T. To further confirm the influence of cooling history on the zero-field resistivity, we also measured $\rho(T)$ by adopting two different cooling processes: (1) the sample is cooled in zero field from 20 K to 1.9 K, and then the $\rho(T)$ is measured upon warming; (2) the sample is cooled in an applied field of 9 T from 20 K to 1.9 K, and the field is then decreased to zero before measuring $\rho(T)$ upon warming. Differences are expected for $\rho(T)$ measured by different processes when there are multiple magnetic domains in the sample, which has been illustrated in a number of compounds, such as ferromagnetic domains in SrRuO$_3$ \cite{SrRuO3} and MnP \cite{MnP}, all-in-all-out typed AFM domains in Nd$_2$Ir$_2$O$_7$ \cite{Nd2Ir2O7}, and competing domains in EuCuSb \cite{EuCuSb}.  In these multi-domain cases, the above-mentioned process (1) was termed as measuring in the \emph{untrained} state (multi-domain state), while process (2) was termed as measured in the \emph{trained} state (domain-aligned state). Our $\rho(T)$ data about EuPtAs is shown in Fig.~\ref{fig7} (a), where a negative magnetoresistance  is observed after the \emph{training} processes, being consistent with that observed in $\rho(H)(H \parallel c)$ shown in Fig.~\ref{fig7} (b). This observation is similar to that in SrRuO$_3$ and MnP with multiple ferromagnetic domains \cite{SrRuO3,MnP}, where the alignment of the magnetic domains along the field direction in the trained state leads to a reduction of the resistivity. \tcr{This is in contrast to that reported for Nd$_2$Ir$_2$O$_7$ and EuCuSb, which both show a positive magnetoresistance} \cite{EuCuSb, Nd2Ir2O7}, where it was proposed that Nd$_2$Ir$_2$O$_7$ hosts metallic domain walls due to the non-trivial topology \cite{Nd2Ir2O7,Nd2Ir2O7_PRB}, while in EuCuSb, the applied field may preferentially select for the domain with lower forming energy but higher resistance \cite{EuCuSb}. As such, the negative magnetoresistance upon applying a training field gives evidence for multiple magnetic domains. While the absence of remnant magnetization in zero-field suggests that these are antiferromagnetic domains, confirmation of their nature requires the microscopic characterization of the magnetic structure.

\subsection{\textit{H}--\textit{T} phase diagrams}
\begin{figure}[!htb]
\centering\includegraphics[width=0.9\columnwidth]{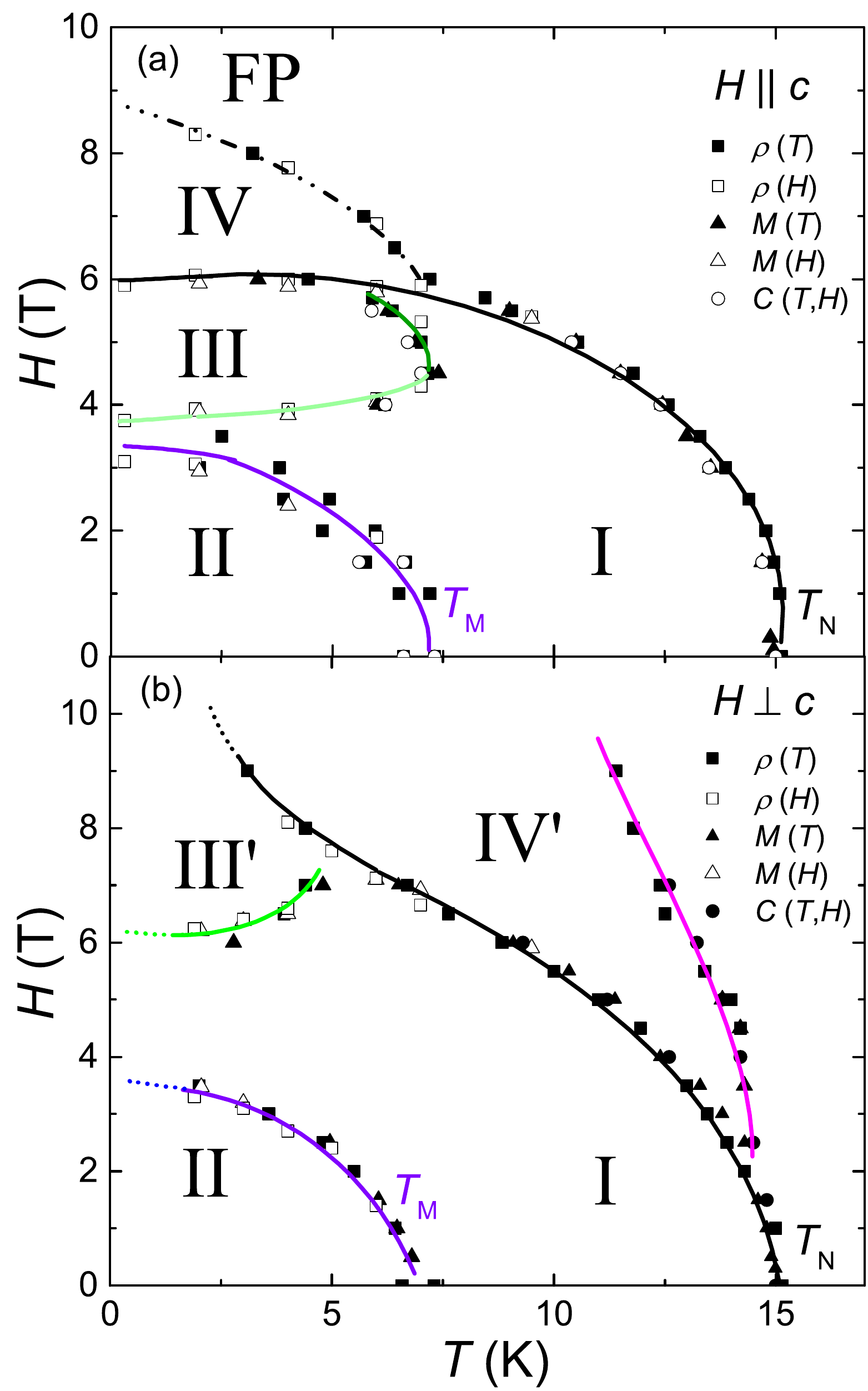}
\caption{(Color online) \emph{T}--\emph{H} phase diagrams for EuPtAs, constructed from measurements of $\rho(T, H)$, \emph{C}(\emph{T}, \emph{H}), and $\chi(T, H)$. The solid lines are guides to the eyes, while the colors correspond to the arrows labelling the transitions in Figs.~\ref{fig4}--\ref{fig6}. The boundary of phase III in (a) is marked with both dark and light green, which correspond to slight differences in the anomalies observed in the respective field ranges. The Roman numerals label the different magnetic phases.}
\label{fig8}
\end{figure}

Based on all the data displayed above, we constructed the \emph{H}--\emph{T} phase diagrams for the two field directions. As shown in Fig.~\ref{fig8}, transition points from different measurements coincide well with each other.
The phase diagrams for \emph{H} $\parallel$ \emph{c} and \emph{H} $\perp$ \emph{c} show some similarities. For example, both \emph{T}$\rm_N$ and \emph{T}$\rm_M$ are monotonically suppressed to lower temperature, and a field-induced phase (phase III and phase III$^{\prime}$) appears below $T\rm_N$. The suppression rates of $T\rm_N$$(H)(H \parallel c)$ and $T\rm_N$$(H)(H \perp c)$ are comparable for $H \leq$ 5 T, suggesting weak magnetocrystalline anisotropy. This is consistent with the small single-ion anisotropy expected for Eu$^{2+}$, where the Hund's rule ground state lacks an orbital moment ($L$ = 0).   

Differences between these two phase diagrams are also apparent.  
For \emph{H} $\parallel$ \emph{c}, $T\rm_N$ is abruptly suppressed at around 6 T, above which a phase \uppercase\expandafter{\romannumeral4} appears at $H \geq$ 6 T located between phase III and the field-polarized (FP) phase, where phase III has a dome shape below $T\rm_N$. 
While for $H \perp c$, $T\rm_N$ is suppressed considerably more slowly for $H \geq$ 6 T,  exhibiting a long tail stretching to higher fields. The field-induced transition (solid green line) appears to terminate upon intersecting $T\rm_N$(\emph{H}) at around 5 K,  which, together with $T\rm_N$(\emph{H}), becomes the boundary of the field-induced phase III$^{\prime}$. On the other hand, at higher fields there is evidence for a field-induced phase IV$^{\prime}$ appearing above $T\rm_N$, and upon cooling for $H \geq$ 7 T, there is a transition directly from this phase to phase III$^{\prime}$. \tcr{The rather complex phase diagrams could be a result of the presence of both Heisenberg-type magnetic exchange interactions and DM interaction, as have been observed in other noncentrosymmetric compounds \cite{MnSi2011, 2017hall, 2015GaV4S8}.}

There are some notable features of the field-induced phases III and phase III$^{\prime}$. In the temperature dependent measurements, upon cooling through the transition into phase III: $\rho(T)$ has a step-like increase under applied fields of 4--6 T, while $\chi(T)$ displays a jump with thermal hysteresis in fields of 4--4.6 T, but a drop with thermal hysteresis in field range of 4.6--6 T, both exhibiting a rather \emph{linear} behavior after cooling across the transition; $C(T)/T$ shows a sharp asymmetric peak without hysteresis, and exhibits a tail on the lower temperature side.  In the field dependent measurements,  $\rho(H)$ shows an abrupt increase at the lower field phase boundary while display a larger drop at the higher field phase boundary with sizable hysteresis at both phase boundaries. $M(H)$ at 2.1 K jumps by approximately 1 $\mu\rm_B$/Eu at the lower boundary, and approximately 2 $\mu\rm_B$/Eu at the higher field boundary, but increases linearly in between.  On the other hand, upon cooling into phase III$^{\prime}$, $\rho(T)$ exhibits a small but sharp drop, while a sharp increase is observed in $\chi(T)$ followed by a flat plateau at 6 T. In $C(T)/T$, there is an asymmetric peak upon entering phase III$^{\prime}$.  $\rho(H)$ undergoes a smaller drop upon crossing the lower boundary of 
phase III$^{\prime}$,  while $M(H)$ shows a sharp change of slope (together with hysteresis) and increases sublinearly inside phase III$^{\prime}$. 

We are not able to determine the magnetic structures of these different phases solely from our present measurements of bulk properties. However, it is of particular interest to compare the phase diagrams of EuPtAs with those of EuPtSi and CeAlGe, both of which crystallize in noncentrosymmetric structures with the latter being isostructural to EuPtAs. EuPtSi shows a phase diagram that is typical for chiral B20 compounds when the field is applied along crystalline [111] direction, where the skyrmion-hosting A-phase arises out of the low field single-\emph{k} helimagnetic phase \cite{2017hall,2018neutron,2019REXS}. Notably the A-phase in EuPtSi has a fully closed phase boundary, existing only in limited temperature- and field- ranges in the equilibrium conditions, although a metastable A-phase can extend down to 60 mK under field-cooling \cite{EPS_new}. While for CeAlGe, the $T\rm_N$ of 4.4 K is continuously suppressed to lower temperature, and two field-induced transitions below $T\rm_N$ are observed, which descend nearly parallel to each other to lower temperature with increasing field, meaning that the boundary of the non-trivial merons/antimeron phase is open on the low temperature side, likely extending to zero temperature \cite{2020CeAlGe}. This is similar to the case for centrosymmetic Gd$_2$PdSi$_3$, where the skyrmion phase also lacks a boundary on the low temperature side \cite{Gd2PdSi3}, here $\rho(H)$ shows abrupt increases and decreases on either side of the skyrmion phase boundary \cite{arxiv-Gd2PdSi3}, which are similar to those observed for phase III in fig.~\ref{fig6}(c). 

In order to microscopically characterize the magnetic structure of the different phases and to look for evidence of topologically nontrivial spin textures (especially , in phase III),  it is both necessary to look for the presence of a topological Hall effect, and to perform neutron diffraction or magnetic x-ray scattering studies, which are currently underway.

\section{Summary}

In summary, we have synthesized single crystals of the noncentrosymmetric Eu-based compound EuPtAs and studied the physical properties by measuring the resistivity, heat capacity, and magnetic susceptibility. In zero-field, two transitions are found with a second-order AFM transition at \emph{T}$\rm_N$ = 14.8 K and a first-order transition at \emph{T}$\rm_M$ = 7.5 K, which were evidenced to be arising from well-localized 4\emph{f} electronic states of Eu$^{2+}$. 

Measurements were performed in applied magnetic fields up to 9 T parallel and perpendicular to the \emph{c}-axis, which allow us to construct the phase diagrams for  both field directions. Our data show that both \emph{T}$\rm_N$ and \emph{T}$\rm_M$ are continuously suppressed by fields for both $H \parallel c$ and $H \perp c$, and there are multiple field-induced transitions for both directions, either above or below $T\rm_N(\emph{H})$, including a dome-shaped phase III for $H \parallel c$. Furthermore, the resistivity measurements under different cooling processes show clear evidences for multiple magnetic domains in the zero-field phase.
Our study shows the presence of complex temperature--field phase diagrams for EuPtAs, with  two zero-field transitions and multiple field-induced phases. In order to characterize the spin-structure and microscopic nature of the magnetism in EuPtAs, it is of particular interest to determine the magnetic structures using neutron or x-ray scattering.

\begin{acknowledgments}
We thank T. Shang for fruitful discussions,  who acknowledges support from the Natural Science Foundation of Shanghai (Grant Nos. 21ZR1420500 and 21JC1402300). This work was supported by the National Key R$\&$D Program of China (No. 2017YFA0303100, No. 2016YFA0300202), the National Natural Science Foundation of China (No. 12034017 and No. 11974306), the Key R$\&$D Program of Zhejiang Province, China (2021C01002), and the Fundamental Research Funds for the Central Universities. W. Xie thanks support from the committee of SCES2019 (Okayama, Japan).
\end{acknowledgments}

\end{document}